\documentclass[aps,nofootinbib,preprint,superscriptaddress]{revtex4}%
\usepackage{hyperref}
\usepackage{amsmath}
\usepackage{amsfonts}
\usepackage{amssymb}
\usepackage{graphicx}
\usepackage{color}%
\providecommand{\U}[1]{\protect\rule{.1in}{.1in}}

\begin{document}
\title{Scattering Equations as the lowest order $K$-identities in the calculation of
Stringy Scaling of Hard String Scattering Amplitudes}
\author{Sheng-Hong Lai}
\email{shlai@cycu.edu.tw}
\affiliation{Department of Physics and Center for High Energy Physics, Chung Yuan Christian
University, Chung Li City, Taiwan, R.O.C.}
\author{Jen-Chi Lee}
\email{jcclee@cc.nctu.edu.tw}
\affiliation{Department of Electrophysics, National Yang Ming Chiao Tung University,
Hsinchu, Taiwan, R.O.C.}
\affiliation{Center for Theoretical and Computational Physics (CTCP), National Yang Ming
Chiao Tung University, Hsinchu, Taiwan, R.O.C.}
\author{Yi Yang}
\email{yangyi3@shanghaitech.edu.cn}
\affiliation{Center for Fundamental Physics (CFP), School of Physical Science and
Technology, ShanghaiTech University, Shanghai, P.R.C.}
\date{\today}

\begin{abstract}
We prove explicitly the $n$-point $K$-identities we proposed previously in the
calculation of one tensor hard string scattering amplitudes ($HSSA$). These
$K$-identities were the key to obtain the stringy scaling behavior in the
saddle point calculation of the $n$-point $HSSA$ and, on the other hand, to
consistently match with the calculation of decoupling of zero norm states.

Moreover, we introduce a $G$ function to generate an infinite set of
generalized $K$-identities ($GKI$). The lowest order set of these $GKI$ is the
scattering equations ($SE$) used in the calculation of field theory amplitudes
in the $CHY$ formalism. The next to leading order set of these $GKI$ is the
$K$-identities used previously in the calculation of one tensor $HSSA$. We
expect that the higher order sets of these $GKI$ can find applications in
calculating other $HSSA$.

\end{abstract}
\maketitle

\setcounter{equation}{0}
\renewcommand{\theequation}{\arabic{section}.\arabic{equation}}
\providecommand{\theHequation}{} \renewcommand{\theHequation}{\arabic{section}.\arabic{equation}}

\section{Introduction}

Recent developments of string scattering amplitudes ($SSA$)
\cite{Rep,Lau,over} have shown that there exist infinite linear relations
among $4$-point hard $SSA$ ($HSSA$) of $26D$ open bosonic string theory
\cite{ChanLee,ChanLee2,CHLTY2,CHLTY1}. More recently, this high energy
behavior of string theory which is not shared by the local quantum field
theory has been extended to the higher $n$-point $HSSA$, and the concept of
the so-called \textit{stringy scaling} \cite{hard, Regge,Komaba} was proposed
by the present authors. Stringy scaling implies that the degree of freedom or
number of kinematic variables reduces for the $n$-point $HSSA$.

Historically, the first stringy scaling behavior was conjectured by Gross
\cite{Gross} for the $4$-point $HSSA$ \cite{GM,GM1,Gross,GrossManes} and later
explicitly proved by Taiwan group in \cite{ChanLee,ChanLee2,CHLTY2,CHLTY1}.
Indeed, all functional forms of $4$-point $HSSA$ on the scattering plane (with
$1$ transverse direction) at each fixed mass level are found to be
proportional to each other with \textit{constant} ratios (independent of the
scattering angle $\phi$, or the deficit of the kinematics variable
dim$\mathcal{M}_{1}=1-0=1$). As another example of stringy scaling
\cite{hard}, it can be shown that the degree of freedom of $6$-point $HSSA$
($5$ tachyons and $1$ tensor) with $3$ transverse directions reduces from $8$
to $2$, and dim$\mathcal{M}_{1}=8-2=6$.

The calculation was soon generalized to the $n$-point ($n\geq4$) $HSSA$ with
$n-2$ tachyons and $2$ high energy tensor states at arbitrary mass levels
\cite{two}. It was found that the degree of stringy scaling dim$\mathcal{M}%
_{2}$ for $HSSA$ with more than $2$ transverse directions decreases comparing
to that of the $1$ tensor $HSSA$ dim$\mathcal{M}_{1}$.

In the saddle point calculation of the $HSSA$, the saddle point of the
integration can be exactly solved only for some lower point $HSSA$
calculation. For the cases of general higher $n$-point saddle point $HSSA$
calculation, the saddle point equations ($SPE$) are in general system of
algebraic equations, and thus is extremely difficult to solve for higher $n$.
So instead of solving the $SPE$, the authors proposed a set of $K$-identities
\cite{hard,Komaba} to simplify the calculation and obtain the degree of
stringy scaling.

The $K$-identities were originally proposed for the calculation of $n$-point
$1$ tensor $HSSA$ \cite{hard, Regge,Komaba}. They are the direct result of the
identification of two different calculation of $HSSA$, namely, the saddle
point calculation and the decoupling of zero norm state ($ZNS$) calculation
\cite{Rep,over}. However, only $K$-identities for the $4$-point $HSSA$ were
analytically proved, other $K$-identities of higher point ($n\geq5$) $HSSA$
calculation were only tested numerically \cite{hard}.

In this letter, we will analytically prove the validity of general $n$-point
$K$-identities. Indeed, we will derive a linear relation between the
$K$-identities and the $SPE$ in Eq.(\ref{equ}). This linear relation is valid
even when $x_{j}$ is out of the saddle point. We will call them the
"off-shell" $K$-identities which contain more information than the "on-shell"
$K$-identities in Eq.(\ref{KI}). On the other hand, one can use string theory
to rederive and prove the "on-shell" $K$-identities by identifying the saddle
point calculation and the decoupling of $ZNS$ calculation of $HSSA$.
Incidentally, we will see that the $K$-identities are crucial to demonstrate
the stringy scaling behavior \cite{hard, Regge,Komaba,two} in the saddle point
calculation of $HSSA$.

Moreover, the existence of the "off-shell" $K$-identities derived in
Eq.(\ref{equ}) motivates us to introduce a $G$ function to generate an
infinite set of generalized $K$-identities ($GKI$). It turns out that the
lowest order set of these $GKI$ is the scattering equations ($SE$) in
Eq.(\ref{chy}) used in the calculation of field theory amplitudes in the $CHY$
formalism \cite{chy1,chy2,chy3,chy4,chy5}. The next to leading order set of
these $GKI$ is the $K$-identities in Eq.(\ref{KI}) used previously in the
calculation of one tensor $HSSA$ \cite{hard,Komaba}. We expect that the higher
order sets of these $GKI$ can find applications in calculating other $HSSA$.

It is interesting to note that, historically, $SE$ was first considered in the
context of string theory by Gross and Mende \cite{GM,GM1} in the calculation
of the $4$-point $HSSA$. Very recently, the general $n$-point $SE$ (or the
$SPE$) was considered by the present authors \cite{hard,Komaba} in the
calculation of the $n$-point $HSSA$.

\setcounter{equation}{0}

\section{The $K$-identities in the calculation of $HSSA$}

Let's begin with a quick review of the $K$-identities. In the calculation of
$n$-point $HSSA$ of open bosonic string theory, we define%
\begin{equation}
f=-\sum_{i>j}\frac{k_{i}\cdot k_{j}}{\Lambda}\ln\left(  x_{i}-x_{j}\right)
\text{, \ }\Lambda=-k_{1}\cdot k_{2},i,j=1,2,\cdots,n, \label{f}%
\end{equation}
where $k_{i}$ and $x_{i}$ are the momentum and worldsheet position of the
$i$-th vertex, and the coordinate-dependent $K_{j}$ vectors as \cite{hard}%
\begin{equation}
K_{j}(x_{1},x_{2},\cdots,x_{n})=-\sum_{i\neq j}\frac{k_{i}}{x_{i}-x_{j}%
}\text{, }j=1,\cdots,n. \label{kj}%
\end{equation}
In the saddle point calculation of $HSSA$, the saddle point $(x_{1}%
,x_{2},x_{3},\cdots,x_{n})$ is defined to be the solution of the system of
saddle point equations ($SPE$)%
\begin{equation}
\frac{\partial}{\partial x_{j}}f(x_{1},x_{2},x_{3},\cdots,x_{n})=0,\text{
}j=1,2,\cdots,n,
\end{equation}
which are, in general, system of algebraic equations and is extremely
difficult to solve for higher $n$. For some cases, one can use the $SL(2,R)$
symmetry to fix $3$ points of $x_{j}$.

In our previous calculation of $HSSA$, we proposed the following
$K$-identities \cite{hard}%
\begin{equation}
\tilde{K}_{j}\cdot\tilde{K}_{j}+2k_{j}\cdot\partial_{j}\tilde{K}_{j}=0,
\label{KI}%
\end{equation}
where $\tilde{K}_{j}$ is the value of $K_{j}$ evaluates at the saddle point.
We will see that Eq.(\ref{KI}) can be proved in the hard scattering limit
$(HSL)$ at the saddle point.

\setcounter{equation}{0}

\section{Proving the $n$-point $K$-identities}

We begin with the $4$-point kinematics in the $HSL$
\begin{align}
k_{1} &  =E\left(  1,-1,0\right)  ,\nonumber\\
k_{2} &  =E\left(  1,+1,0\right)  ,\nonumber\\
k_{3} &  =E\left(  -1,-\cos\phi,-\sin\phi\right)  ,\nonumber\\
k_{4} &  =E\left(  -1,\cos\phi,\sin\phi\right)  .
\end{align}
In the calculation of $4$-point $HSSA$, after fixing the $SL(2,R)$ symmetry,
there is only one $SPE$ which is linear, and the saddle point can be easily
solved to be%
\begin{equation}
\tilde{x}_{2}=\frac{1}{1-\tau}\text{, \ }\tau=-\frac{t}{s}=\sin^{2}\frac{\phi
}{2}.
\end{equation}
For the $4$-point $1$ tensor $HSSA$, one easily shows%
\begin{equation}
\tilde{K}_{2}\cdot\tilde{K}_{2}+2k_{2}\cdot\partial_{2}\tilde{K}%
_{2}=0,\label{aaa}%
\end{equation}
where%
\begin{equation}
K_{2}\left(  x_{2}\right)  =\frac{k_{1}}{x_{2}}-\frac{k_{3}}{1-x_{2}}.
\end{equation}
For the cases of higher point $SPE$ with $n\geq5$, it is nontrivial to solve
the system of higher order algebraic equations. So instead of solving the
equations, we will use the $SPE$ directly to prove the $K$-identities. The
corresponding $SPE$ of the $n$-point $HSSA$ are
\begin{equation}
\mathcal{E}_{k}=k_{k}\cdot K_{k}=-\sum_{i\neq k}^{n}\frac{k_{ik}}{x_{i}-x_{k}%
}=0,\qquad k=1,\ldots,n.\label{spt}%
\end{equation}
(Eq.(\ref{spt}) is also known as the $SE$ in the $CHY$ formalism. See section
VI) The coordinate dependent $K_{k}$ vectors are defined in Eq.(\ref{kj}). For
each $k$, we also define $I_{k}$ to be%
\begin{equation}
I_{k}:=K_{k}\cdot K_{k}+2k_{k}\cdot\partial_{k}K_{k}.
\end{equation}
In the $HSL$, $k_{ij}$\ ($i\neq j$) goes to infinity and $k_{j}^{2}$ is finite
and can be ignored, one can calculate
\begin{equation}
I_{k}=\sum_{i\neq k,j\neq k,i>j}^{n}\frac{2k_{ij}}{\left(  x_{i}-x_{k}\right)
\left(  x_{j}-x_{k}\right)  }-\sum_{i\neq k}^{n}\frac{2k_{ik}}{\left(
x_{i}-x_{k}\right)  ^{2}}.\label{nkk}%
\end{equation}
For each pair with $i>j$, we apply the partial-fraction decomposition
\begin{equation}
\frac{1}{\left(  x_{i}-x_{k}\right)  \left(  x_{j}-x_{k}\right)  }=\frac
{1}{\left(  x_{i}-x_{k}\right)  \left(  x_{j}-x_{i}\right)  }+\frac{1}{\left(
x_{j}-x_{k}\right)  \left(  x_{i}-x_{j}\right)  },
\end{equation}
and Eq.(\ref{nkk}) becomes%
\begin{equation}
I_{k}=\sum_{i\neq k,j\neq k,i>j}^{n}\left[  \frac{2k_{ij}}{\left(  x_{i}%
-x_{k}\right)  \left(  x_{j}-x_{i}\right)  }+\frac{2k_{ij}}{\left(
x_{j}-x_{k}\right)  \left(  x_{i}-x_{j}\right)  }\right]  -\sum_{i\neq k}%
^{n}\frac{2k_{ik}}{\left(  x_{i}-x_{k}\right)  ^{2}}.\label{ii}%
\end{equation}
After some manipulation, Eq.(\ref{ii}) can be written as
\begin{equation}
I_{k}=\sum_{i,j\neq k,i\neq j}^{n}\frac{2k_{ij}}{\left(  x_{k}-x_{i}\right)
\left(  x_{i}-x_{j}\right)  }-\sum_{i\neq k}^{n}\frac{2k_{ik}}{\left(
x_{i}-x_{k}\right)  ^{2}}.\label{iii}%
\end{equation}
Note that the term in the last sum of the above equation can be rewritten as
\begin{equation}
-\frac{2k_{ik}}{\left(  x_{i}-x_{k}\right)  ^{2}}=\frac{2}{x_{i}-x_{k}}%
\frac{k_{ik}}{x_{k}-x_{i}}.
\end{equation}
So one can rewrite Eq.(\ref{iii}) as
\begin{equation}
I_{k}=\sum_{i\neq k}^{n}\frac{2}{x_{k}-x_{i}}\left[  \sum_{j\neq i,j\neq
k}^{n}\frac{k_{ij}}{x_{i}-x_{j}}+\frac{k_{ik}}{x_{i}-x_{k}}\right]  ,
\end{equation}
which gives%
\begin{equation}
I_{k}=\sum_{i\neq k}^{n}\frac{2}{x_{k}-x_{i}}\left[  \sum_{j\neq i}^{n}%
\frac{k_{ij}}{x_{i}-x_{j}}\right]  .\label{iik}%
\end{equation}
The expression in the square bracket of Eq.(\ref{iik}) is exactly the left
hand side of the $SP$ $\mathcal{E}_{i}$ in Eq.(\ref{spt}). Thus the whole
calculation reduces to the linear relation
\begin{equation}
I_{k}=\sum_{i\neq k}^{n}\frac{2}{x_{k}-x_{i}}\mathcal{E}_{i}.\label{equ}%
\end{equation}
We thus have derived Eq.(\ref{equ}) in the $HSL$. Since $\mathcal{E}_{i}$
vanishes at the saddle point, Eq.(\ref{equ}) immediately implies $I_{k}=0$ for
each $k=1,\ldots,n$. For illustration, we give an example of the linear
relation in Eq.(\ref{equ}) for the case of $n=5$ in the matrix form
\begin{equation}%
\begin{bmatrix}
I_{1}\\
I_{2}\\
I_{3}\\
I_{4}\\
I_{5}%
\end{bmatrix}
=-2%
\begin{bmatrix}
0 & \frac{1}{x_{2}-x_{1}} & \frac{1}{x_{3}-x_{1}} & \frac{1}{x_{4}-x_{1}} &
\frac{1}{x_{5}-x_{1}}\\
\frac{1}{x_{1}-x_{2}} & 0 & \frac{1}{x_{3}-x_{2}} & \frac{1}{x_{4}-x_{2}} &
\frac{1}{x_{5}-x_{2}}\\
\frac{1}{x_{1}-x_{3}} & \frac{1}{x_{2}-x_{3}} & 0 & \frac{1}{x_{4}-x_{3}} &
\frac{1}{x_{5}-x_{3}}\\
\frac{1}{x_{1}-x_{4}} & \frac{1}{x_{2}-x_{4}} & \frac{1}{x_{3}-x_{4}} & 0 &
\frac{1}{x_{5}-x_{4}}\\
\frac{1}{x_{1}-x_{5}} & \frac{1}{x_{2}-x_{5}} & \frac{1}{x_{3}-x_{5}} &
\frac{1}{x_{4}-x_{5}} & 0
\end{bmatrix}%
\begin{bmatrix}
\mathcal{E}_{1}\\
\mathcal{E}_{2}\\
\mathcal{E}_{3}\\
\mathcal{E}_{4}\\
\mathcal{E}_{5}%
\end{bmatrix}
.
\end{equation}

\setcounter{equation}{0}

\section{Deriving $K$-identities from string theory}

In this section, we will see that the $K$-identities proved in the previous
section can be remarkably rederived from string theory! Indeed, we will derive
and "prove" the $K$-identities from the calculation of $HSSA$ in string theory
\cite{hard}. For simplicity, we will also use an example to illustrate the
calculation. Note first that for $n$-point $HSSA$ with $n\geq5$, the
scattering process may not be on the scattering plane with only $r=1$
transverse direction as for the case of $n=4$. Indeed, for $n\geq4$ the
general high energy states at each fixed mass level $M^{2}=2(N-1)$ can be
written as\bigskip\ \cite{hard}%
\begin{equation}
\left\vert \left\{  p_{i}\right\}  ,2m,2q\right\rangle =\left(  \alpha
_{-1}^{T_{1}}\right)  ^{N+p_{1}}\left(  \alpha_{-1}^{T_{2}}\right)  ^{p_{2}%
}\cdots\left(  \alpha_{-1}^{T_{r}}\right)  ^{p_{r}}\left(  \alpha_{-1}%
^{L}\right)  ^{2m}\left(  \alpha_{-2}^{L}\right)  ^{q}\left\vert
0;k\right\rangle ,
\end{equation}
where $\sum_{i=1}^{r}p_{i}=-2(m+q)$ with\ $r\leq24$, and $p_{j}$, $m$ and $q$
are nonnegative integers.\ One generalizes the transverse polarization
$e^{T}=(0,0,1)$ on the scattering plane in the $4$-point $HSSA$ calculation to
$e^{\hat{T}}=(0,0,\vec{\omega})$ where%
\begin{equation}
\omega_{i}=\cos\theta_{i}\prod\limits_{\sigma=1}^{i-1}\sin\theta_{\sigma
}\text{, }i=1,\cdots,r,\text{ }\theta_{r}=0, \label{228}%
\end{equation}
are the solid angles in the transverse space spanned by $24$ transverse
directions $e^{T_{i}}$. The ratios among $HSSA$ for different string states
(of say vertices $v_{2}$) at each fixed mass level can then be calculated by
the method of decoupling of $ZNS$ to be \cite{hard}%
\begin{equation}
\frac{\mathcal{T}^{\left(  \left\{  p_{i}\right\}  ,2m,2q\right)  }%
}{\mathcal{T}^{\left(  \left\{  0_{i}\right\}  ,0,0\right)  }}=\frac{\left(
2m\right)  !}{m!}\left(  \frac{-1}{2M}\right)  ^{2m+q}\prod_{i=1}^{r}%
\omega_{i}^{p_{i}}, \label{100}%
\end{equation}
where $\mathcal{T}^{\left(  \left\{  0_{i}\right\}  ,0,0\right)  }$ is the
$HSSA$ of leading Regge trajectory state at mass level $M^{2}=2(N-1)$.\ These
ratios are valid to all string loop orders. Note that vertices $v_{1}$,
$v_{3}$, $\cdots$,$v_{n}$ in the $n$-point $HSSA$ $\mathcal{T}^{\left(
\left\{  p_{i}\right\}  ,2m,2q\right)  }$ in Eq.(\ref{100}) can be any string
states, and we have omitted their indices.

On the other hand, one expects to reproduce Eq.(\ref{100}) from the saddle
point calculation of $HSSA$. In the $HSL$, as $p=E\rightarrow\infty$, we
define the $26$-dimensional momenta in the CM frame to be\ \cite{hard}%
\begin{align}
k_{1}  &  =E\left(  1,-1,0^{r}\right)  ,\nonumber\\
k_{2}  &  =E\left(  1,+1,0^{r}\right)  ,\nonumber\\
&  \vdots\nonumber\\
k_{j}  &  =-q_{j}\left(  1,\Omega_{1}^{j},\cdots,\Omega_{r}^{j},\Omega
_{r+1}^{j}\right)  ,
\end{align}
where $j=3,4,\cdots,n$, and%
\begin{equation}
\Omega_{i}^{j}=\cos\phi_{i}^{j}\prod\limits_{\sigma=1}^{i-1}\sin\phi_{\sigma
}^{j}\text{ with }\phi_{j-1}^{j}=0,\text{ }\phi_{i>r}^{j}=0\text{ and }%
r\leq\min\left\{  n-3,24\right\}  ,
\end{equation}
are the solid angles in the $\left(  j-2\right)  $-dimensional spherical space
with $\sum_{i=1}^{j-2}\left(  \Omega_{i}^{j}\right)  ^{2}=1$.

Since the ratios in Eq.(\ref{100}) are independent of the vertices $v_{1}$,
$v_{3}$, $\cdots$,$v_{n}$, for simplicity, we choose them to be tachyons. The
one tensor (with mass $M_{2}^{2}=2(N-1)$ at the second vertex) $n$-point
$HSSA$ of open bosonic string with $r$ transverse directions can be calculated
by using saddle point method to be \cite{hard} ($K_{2}^{T_{j}}=K_{2}\cdot
e^{T_{j}}$)%

\begin{align}
\mathcal{T}^{\left(  \left\{  p_{i}\right\}  ,2m,2q\right)  }  & =2\sqrt{\pi
}e^{-\Lambda\tilde{f}}\left\vert \tilde{K}_{2}\right\vert ^{N-1}\left(
\prod_{i=3}^{n-2}\left(  \frac{\tilde{x}_{i}}{\tilde{x}_{i+1}}\right)
^{i-2-N}\right)  \nonumber\\
& \cdot\frac{\left(  2m\right)  !}{m!}\left(  \frac{-1}{2M_{2}}\right)
^{2m+q}\prod\limits_{j=1}^{r}\left(  \frac{\tilde{K}_{2}^{T_{j}}}%
{\sqrt{2\Lambda\tilde{f}_{22}}}\right)  ^{p_{j}},\label{rat}%
\end{align}
which leads to the ratios%

\begin{equation}
\frac{\mathcal{T}^{\left(  \left\{  p_{i}\right\}  ,2m,2q\right)  }%
}{\mathcal{T}^{\left(  \left\{  0_{i}\right\}  ,0,0\right)  }}=\frac{\left(
2m\right)  !}{m!}\left(  \frac{-1}{2M_{2}}\right)  ^{2m+q}\prod\limits_{j=1}%
^{r}\left(  \frac{\tilde{K}_{2}^{T_{j}}}{\sqrt{2\Lambda\tilde{f}_{22}}%
}\right)  ^{p_{j}}. \label{zn}%
\end{equation}
In Eq.(\ref{rat}), $\tilde{K}_{2}$ means $K_{2}$ evaluated at the saddle-point
etc. and $f_{22}=\frac{\partial^{2}f}{\partial x_{2}^{2}}$ with $f$ \ defined
in Eq.(\ref{f}). Finally, the identification of Eq.(\ref{zn}) and
Eq.(\ref{100}) gives \cite{hard} the $K$-identity in Eq.(\ref{KI}) for the
case of $j=2$.

As an illustration, for the example of $n=6$ and $r=3$ with $5$ tachyons and
$1$ tensor (at $x_{2}$) $HSSA$ calculation, Eq.(\ref{228}) gives
\begin{equation}
\vec{\omega}=(\omega_{1},\omega_{2},\omega_{3})=(\cos\theta_{1},\sin\theta
_{1}\cos\theta_{2},\sin\theta_{1}\sin\theta_{2})\text{.}%
\end{equation}
Eq.(\ref{zn}) and Eq.(\ref{100}) can be identified for any $p_{2}$ and $p_{3}$
if%
\begin{equation}
\tilde{K}_{2}^{T_{j}}=\sqrt{2\Lambda\tilde{f}_{22}}\text{ }\omega_{j}\text{,
}j=1,\cdots,r, \label{c1}%
\end{equation}
which allow us to express $\theta_{1}$ and $\theta_{2}$ in terms of kinematics
variables as \cite{hard}%
\begin{equation}
\theta_{1}=\arctan\frac{\sqrt{\left(  \tilde{K}_{2}^{T_{2}}\right)
^{2}+\left(  \tilde{K}_{2}^{T_{3}}\right)  ^{2}}}{\tilde{K}_{2}^{T_{1}}%
}\text{, }\theta_{2}=\arctan\frac{\tilde{K}_{2}^{T_{3}}}{\tilde{K}_{2}^{T_{2}%
}}. \label{scaling}%
\end{equation}
So the ratios of $6$-point $HSSA$ with $r=3$ depend only on $2$ kinematics
variables $\theta_{1}$ and $\theta_{2}$ instead of $8$. Therefore, the deficit
of the kinematic variables or the degree of stringy scaling is dim$\mathcal{M}%
_{1}=8-2=6$. Note that $\theta_{1}$ and $\theta_{2}$ can be expressed in terms
of $8$ kinematics variables if one can solve the saddle points $\tilde{x}_{j}$.

We see that Eq.(\ref{c1}) gives%
\begin{equation}
\left(  \text{ }\tilde{K}_{2}^{T_{1}}\right)  ^{2}+\left(  \tilde{K}%
_{2}^{T_{2}}\right)  ^{2}+\left(  \text{ }\tilde{K}_{2}^{T_{3}}\right)
^{2}=2\Lambda\tilde{f}_{22}, \label{sk}%
\end{equation}
which can be written as the $K$-identity in Eq.(\ref{KI}) for the case of
$n=6$, $j=2$ and $r=3$.

For the general\textbf{ }$n$\textbf{-}point\textbf{ }$1$\textbf{
}tensor\textbf{ }$HSSA$, it can be shown that\ the degree of stringy scaling
is \cite{hard}
\begin{equation}
\text{dim}\mathcal{M}_{1}=\frac{\left(  r+1\right)  \left(  2n-r-6\right)
}{2}.
\end{equation}
\setcounter{equation}{0}

\section{Scattering Equations as the lowest order $K$-identities}

It is important to note that the relation%
\begin{equation}
I_{k}:=K_{k}\cdot K_{k}+2k_{k}\cdot\partial_{k}K_{k}=\sum_{i\neq k}%
\frac{2\mathcal{E}_{i}}{x_{k}-x_{i}}, \label{ok}%
\end{equation}
calculated in Eq.(\ref{equ}) is valid even when $x_{k}$ is out of the saddle
point! Indeed, it is also valid if we replace the real variable $x_{k}$ by a
complex variable $z_{k}$ and thus extends the M\"{o}bius symmetry to
$SL(2,C)$. This seems to suggest that the information contained in the
relation in Eq.(\ref{ok}) is much more than the $K$-identities $I_{k}=0$
themselves which are valid only at the saddle point satisfying the $SPE$ in
Eq.(\ref{spt}). This motivates us to define the following $G(z)$ function%
\begin{equation}
G(z,z_{1},\cdots,z_{n})=G(z)=\sum_{i=1}^{n}\frac{2\mathcal{E}_{i}}{z-z_{i}}.
\label{g}%
\end{equation}
On the other hand, we want to understand how to reproduce the "on-shell"
$K$-identities from the "off-shell" $G(z)$ function. In view of the first
$K_{k}^{2}{}$ term in the relation in Eq.(\ref{ok}), let's define the
following "off-shell" $K(z)$ function%
\begin{equation}
K(z,z_{1},\cdots,z_{n})=K(z)=\sum_{i=1}^{n}\frac{k_{i}}{z-z_{i}}, \label{k}%
\end{equation}
and calculate $K(z)^{2}$. After doing some algebra, we obtain%
\begin{equation}
K(z)^{2}=\sum_{i}\frac{k_{i}^{2}}{(z-z_{i})^{2}}+\sum_{i=1}^{n}\frac
{2\mathcal{E}_{i}}{z-z_{i}}. \label{k2}%
\end{equation}
In the $HSL$, the first term above can be ignored and we get%
\begin{equation}
K(z)^{2}=\sum_{i=1}^{n}\frac{2}{z-z_{i}}\mathcal{E}_{i}=G(z)\text{ }(HSL).
\label{sad}%
\end{equation}
So we expect to reproduce the "on-shell" $K$-identities Eq.(\ref{KI}) from the
more general "off-shell" $G(z)$ function or $K(z)^{2}$ function in the $HSL$.
Let's begin with the Laurent expansion of $K(z)$ around the point $z_{i}$. For
a fixed point $z_{i}$, let's define
\begin{equation}
\omega=z-z_{i}.
\end{equation}
The $K$-function in Eq.(\ref{k}) can be separated into a singular part and a
regular part:
\begin{equation}
K(z)=\sum_{j}\frac{k_{j}}{z-z_{j}}=\frac{k_{i}}{\omega}+\sum_{j\neq i}%
\frac{k_{j}}{z_{i}-z_{j}+\omega}.
\end{equation}
For the regular part of $K(z)$, we can use the geometric series to get%
\begin{equation}
\frac{1}{z_{i}-z_{j}+\omega}=\frac{1}{z_{i}-z_{j}}\frac{1}{1+\omega
/(z_{i}-z_{j})}=\sum_{r=0}^{\infty}\frac{(-1)^{r}\omega^{r}}{(z_{i}%
-z_{j})^{r+1}}.
\end{equation}
On the other hand, from the definition of $K_{k}$ in Eq.(\ref{kj}), we can
calculate%
\begin{equation}
\partial_{i}^{r}K_{i}=(-1)^{r}r!\sum_{j\neq i}\frac{k_{j}}{(z_{i}-z_{j}%
)^{r+1}}.
\end{equation}
Therefore
\begin{equation}
K(z)=\frac{k_{i}}{\omega}+\sum_{r=0}^{\infty}\frac{\omega^{r}}{r!}\partial
_{i}^{r}K_{i}=\frac{k_{i}}{\omega}+e^{\omega\partial_{i}}K_{i}. \label{exp}%
\end{equation}
The first five terms of Eq.(\ref{exp}) are
\begin{equation}
K(z)=\frac{k_{i}}{\omega}+K_{i}+\omega\partial_{i}K_{i}+\frac{\omega^{2}}%
{2}\partial_{i}^{2}K_{i}+\frac{\omega^{3}}{6}\partial_{i}^{3}K_{i}%
+O(\omega^{4}). \label{K-expand}%
\end{equation}
The next step is to calculate $K(z)^{2}$. One easily obtains the following
"off-shell" expansion%
\begin{align}
K(z)^{2}  &  =\left(  \frac{k_{i}}{\omega}+K_{i}+\omega\partial_{i}K_{i}%
+\frac{\omega^{2}}{2}\partial_{i}^{2}K_{i}+\frac{\omega^{3}}{6}\partial
_{i}^{3}K_{i}+O(\omega^{4})\right)  ^{2}\nonumber\\
&  =\frac{k_{i}^{2}}{\omega^{2}}+\frac{2k_{i}\cdot K_{i}}{\omega}+\left(
K_{i}^{2}+2k_{i}\cdot\partial_{i}K_{i}\right) \nonumber\\
&  +\omega\left[  2K_{i}\cdot\partial_{i}K_{i}+k_{i}\cdot\partial_{i}^{2}%
K_{i}\right] \nonumber\\
&  +\omega^{2}\left[  (\partial_{i}K_{i})^{2}+K_{i}\cdot\partial_{i}^{2}%
K_{i}+\frac{1}{3}k_{i}\cdot\partial_{i}^{3}K_{i}\right]  +O(\omega^{3}).
\label{666}%
\end{align}
Since $K(z)^{2}\equiv0$ in the $HSL$ at the saddle point by Eq.(\ref{sad}),
each coefficient in the expansion must vanish. We thus obtain infinite number
of \textit{on-shell} $K$-identities. The first few identities are
\begin{align}
k_{i}^{2}  &  =0,\label{11}\\
k_{i}\cdot\tilde{K}_{i}  &  =0,\label{22}\\
\tilde{K}_{i}\cdot\tilde{K}_{i}+2k_{i}\cdot\partial_{i}\tilde{K}_{i}  &
=0,\label{33}\\
2\tilde{K}_{i}\cdot\partial_{i}\tilde{K}_{i}+k_{i}\cdot\partial_{i}^{2}%
\tilde{K}_{i}  &  =0,\label{44}\\
(\partial_{i}\tilde{K}_{i})^{2}+\tilde{K}_{i}\cdot\partial_{i}^{2}\tilde
{K}_{i}+\frac{1}{3}k_{i}\cdot\partial_{i}^{3}\tilde{K}_{i}  &  =0. \label{55}%
\end{align}

Eq.(\ref{11}) is the massless condition in the $CHY$ formalism, which
corresponds to the $HSL$ condition in the $HSSA$ calculation. We see that
Eq.(\ref{22}) is the $SE$ or $SPE$ in Eq.(\ref{spt}). See also Eq.(\ref{chy})
in sectionVI. On the other hand, Eq.(\ref{33}) is the celebrated "on-shell"
$K$-identities we obtained in Eq.(\ref{KI}) previously in the calculation of
$HSSA$. In addition, we have derived infinite number of identities for $K_{i}%
$. Indeed, for $m\geq0$, one easily calculates%
\begin{equation}
\frac{2}{\left(  m+1\right)  !}k_{i}\cdot\partial_{i}^{m+1}\tilde{K}_{i}%
+\frac{1}{m!}\partial_{i}^{m}\left(  \tilde{K}_{i}\cdot\tilde{K}_{i}\right)
=0. \label{66}%
\end{equation}
If we consider the $SE$ in Eq.(\ref{22}) as an algebraic equation solved by
the saddle points $(\tilde{z}_{1},\cdots,\tilde{z}_{n})$. Other higher
$K$-identities in Eq.(\ref{33}) to Eq.(\ref{66}) correspond to higher order
algebraic equations solved by the same saddle points $(\tilde{z}_{1}%
,\cdots,\tilde{z}_{n})$.

So from this point of view, $SE$\textit{ can be thought of as the lowest order
set of infinite number of }$K$\textit{-identities. }While Eq.(\ref{22}) was
useful in calculating field theory amplitudes as pointed out in $CHY$
formalism, Eq.(\ref{33}) was important in calculating $HSSA$ first pointed out
in \cite{hard,Regge, Komaba}. We expect that the higher order sets of these
$GKI$ in Eq.(\ref{44}) to Eq.(\ref{66}) may find applications in calculating
other $HSSA$.

One can do similar expansion for $G(z)$. By using Eq.(\ref{g}), we easily get%
\begin{equation}
G(z)=\frac{2\mathcal{E}_{i}}{\omega}+\sum_{r=0}^{\infty}\sum_{j\neq i}%
\frac{2\mathcal{E}_{j}(-1)^{r}\omega^{r}}{(z_{i}-z_{j})^{r+1}}. \label{gg}%
\end{equation}
By identifying Eq.(\ref{gg}) and Eq.(\ref{666}), we obtain an infinite number
of \textit{off-shell} $K$-identities.

For the first term, we get%
\begin{equation}
\mathcal{E}_{i}=-\sum_{j\neq i}\frac{k_{ij}}{z_{j}-z_{i}}. \label{se}%
\end{equation}
For $r=0$, we reproduce the previous off-shell $K$-identities we obtained in
Eq.(\ref{equ})%
\begin{equation}
I_{i}=K_{i}\cdot K_{i}+2k_{i}\cdot\partial_{i}K_{i}=-\sum_{j\neq i}\frac
{2}{z_{j}-z_{i}}\mathcal{E}_{j}. \label{kk}%
\end{equation}
Note that Eq.(\ref{kk}) gives Eq.(\ref{33}) at the saddle point. For $r=1$, we
obtain%
\begin{equation}
2K_{i}\cdot\partial_{i}K_{i}+k_{i}\cdot\partial_{i}^{2}K_{i}=-\sum_{j\neq
i}\frac{2}{(z_{j}-z_{i})^{2}}\mathcal{E}_{j}, \label{k3}%
\end{equation}
and for $r=2$, we get%
\begin{equation}
(\partial_{i}K_{i})^{2}+K_{i}\cdot\partial_{i}^{2}K_{i}+\frac{1}{3}k_{i}%
\cdot\partial_{i}^{3}K_{i}=-\sum_{j\neq i}\frac{2}{(z_{j}-z_{i})^{3}%
}\mathcal{E}_{j}. \label{k4}%
\end{equation}
In general for $r=m$, we obtain
\begin{equation}
\frac{2}{\left(  m+1\right)  !}k_{i}\cdot\partial_{i}^{m+1}K_{i}+\frac{1}%
{m!}\partial_{i}^{m}\left(  K_{i}\cdot K_{i}\right)  =-\sum_{j\neq i}\frac
{2}{(z_{j}-z_{i})^{m+1}}\mathcal{E}_{j}. \label{k5}%
\end{equation}
In particular, if we put the "on-shell" conditions, $\mathcal{E}_{i}=0$, on
the above "off-shell" equations, we reproduce the previous "on-shell"
$K$-identities. Note that these off-shell $K$-identities are valid even when
$z_{j}$ are out of the saddle point.

\setcounter{equation}{0}

\section{The $K$-identities and $CHY$ formalism}

In the calculation of field theory scattering amplitudes of $CHY$ formalism,
$CHY$ proposed the following formula \cite{chy1,chy2,chy3,chy4,chy5}%
\begin{equation}
\mathcal{A}_{n}=\int\frac{d^{n}z_{i}}{\text{Vol }[SL(2,C)]}\left[
{\displaystyle\prod}
\delta(\mathcal{E}_{i})\right]  \text{ }\mathcal{I}=\int d\mu_{n}\text{
}\mathcal{I}\text{.} \label{chy}%
\end{equation}
See \cite{fur1,fur2,fur3,fur4,fur5,fur6} for further developments. In the
above proposal, the measure $d\mu_{n}$ and, in particular, the $SE$ $%
{\displaystyle\prod}
\delta(\mathcal{E}_{i})$ (or $SPE$) are universal to all field theories of
massless particles (see Eq.(\ref{spt}) and Eq.(\ref{22})), while the integrand
$\mathcal{I}$ contains the specific information of the theory, such as YM
gauge field theory.

$CHY$ formalism provides a compact way to calculate tree-level scattering
amplitudes in field theory by reducing the amplitude integration to a discrete
algebraic sum over the solutions of the scattering equations ($SE$). Among $n$
scattering equations only $(n-3)$ of them are independent due to the $SL(2,C)$
M\"{o}bius symmetry of the Riemann sphere $CP^{1}$. The number of solutions of
the scattering equations is $(n-3)!$ by Bezout's theorem which, for the YM
case, matches with the number of independent color-stripped scattering amplitudes.

There are at least two approaches to understand the $SE$ and the origin of
$CHY$ formalism. The first one is through the ambitwistor string
\cite{mason1,mason2,mason3, Stieberger} or worldsheet approach and the second
is the $HSSA$ approach.

\subsection{Worldsheet approach}

For this approach one ingredient is to consider the connection between the map
from closed string worldsheet $CP^{1}$ with $n$-puncture to the null cone and
the scattering data of $n$ massless particles. The map is given by%
\begin{equation}
k_{a}^{\mu}=%
{\displaystyle\oint\limits_{\left\vert \sigma-\sigma_{a}\right\vert =\epsilon
}}
d\sigma P^{\mu}(\sigma),
\end{equation}
where%
\begin{equation}
P^{\mu}(\sigma)=\sum_{a=1}^{n}\frac{k_{a}^{\mu}}{\sigma-\sigma_{a}}. \label{p}%
\end{equation}
Note that Eq.(\ref{p}) is exactly the same with Eq.(\ref{k}) we defined
previously. The next step is to require $P^{\mu}$ to live in the null cone%
\begin{equation}
P^{\mu}(\sigma):CP^{1}\rightarrow\text{ null cone,}%
\end{equation}
with%
\begin{equation}
P^{2}(\sigma)=\sum_{a,b}^{n}\frac{k_{a}\cdot k_{b}}{(\sigma-\sigma_{a}%
)(\sigma-\sigma_{b})}=0. \label{p2}%
\end{equation}
We note that Eq.(\ref{p2}) is the same as Eq.(\ref{k2}) we calculated
previously. Finally, by using Eq.(\ref{sad}), one gets
\begin{equation}
\mathcal{E}_{i}=-\sum_{j\neq i}\frac{k_{ij}}{\sigma_{j}-\sigma_{i}}=0,
\label{sadd2}%
\end{equation}
which is the $SE$ in Eq.(\ref{chy}). Note that the $SE$ in Eq.(\ref{sadd2}) is
exactly the same with the $SPE$ in Eq.(\ref{spt}).

\subsection{$HSSA$ approach}

Instead of calculating field theory scattering amplitudes in Eq.(\ref{chy}),
an alternative and more direct approach to obtain $SE$ is in the calculation
of $n$-point $HSSA$. In this point of view, $SE$ arises naturally as the $SPE$
in the calculation of $HSSA$ \cite{hard, Regge,Komaba}. For the $4$-point
case, see \cite{GM,GM1}. However, in contrast to solving the algebraic $SE$ in
the field theory amplitude calculation, in the $HSSA$ calculation we proposed
a set of the $K$-identities \cite{hard, Regge,Komaba} in Eq.(\ref{kk})
(without solving the saddle points) to obtain infinite ratios among $HSSA$ at
each fixed mass level of the string spectrum.

Moreover, there are two important ingredients in the calculation of $HSSA$
which are not shared by the $CHY$ field theory calculation. The first one is
the introduction of the $G$ function in Eq.(\ref{g}) to generate an infinite
set of $GKI$ in Eq.(\ref{22}) to Eq.(\ref{66}) which are satisfied by the
saddle points $(\tilde{z}_{1},\cdots,\tilde{z}_{n})$. The lowest order set of
these $GKI$ is the $SE$ in Eq.(\ref{22}) used in the calculation of field
theory amplitude in the $CHY$ formalism. The next to leading order set of
these $GKI$ is the $K$-identities in Eq.(\ref{33}) used previously in the
calculation of one tensor $HSSA$ \cite{hard, Regge,Komaba}.

The second one is the \textit{stringy scaling} \cite{hard, Regge,Komaba}
behavior of $HSSA$ in Eq.(\ref{scaling}) which is closely related to the
$K$-identities in Eq.(\ref{33}). We expect that the higher order sets of these
$GKI$ in Eq.(\ref{44}) to Eq.(\ref{66}) may find applications in calculating
other $HSSA$.

\setcounter{equation}{0}

\begin{acknowledgments}
We thank Y. Okawa, T. Okuda and T. Yoneya for discussions. Sheng-Hong Lai is
supported by the National Science and Technology Council (NSTC), Taiwan, under
Grant No. NSTC 114-2112-M-033 -014 -MY3. Jen-Chi Lee is supported by a Grant
from NYCU. Yi Yang is supported by the start-up funding 2025F0201-000-02 and
the CFP funding 2026A0201-405-09 by ShanghaiTech University.
\end{acknowledgments}


\end{document}